\documentstyle[epsfig]{mn}

\newcommand{\reference}{\bibitem}

\title[]{Failed gamma-ray bursts and orphan afterglows}

\author[]{Y. F. Huang,$^{1,2 \; \star}$ Z. G. Dai$^{1 \; \star}$ 
   and T. Lu$^{1,2}$ 
\thanks{E-mail: hyf@nju.edu.cn(YFH); daizigao@public1.ptt.js.cn(ZGD);  
        tlu@nju.edu.cn(TL)} \\
$^1${\sl Department of Astronomy, Nanjing University, Nanjing 210093, 
         P. R. China} \\
$^2${\sl LCRHEA, Institute for High-Energy Physics, Chinese Academy of 
         Sciences, Beijing 100039, P. R. China} } 
\date{Accepted ......  Received ......; in original form ......  
      (MNRAS, 2002, in press) }

\begin{document}
\voffset=-0.5 in

\maketitle
\begin{abstract}
It is believed that orphan afterglow searches can help to measure the beaming
angle in gamma-ray bursts (GRBs). Great expectations have been put on this
method. We point out that the method is in fact not as simple as we
originally expected. Due to the baryon-rich environment that is common
to almost all popular progenitor models, there should be many failed
gamma-ray bursts, i.e., fireballs with Lorentz factor much less than
100 --- 1000, but still much larger than unity. In fact, the number of failed
gamma-ray bursts may even be much larger than that of successful bursts. Owing
to the existence of these failed gamma-ray bursts, there should be many
orphan afterglows even if GRBs are due to isotropic fireballs, then the
simple discovery of orphan afterglows never means that GRBs be
collimated. Unfortunately, to distinguish a failed-GRB orphan and a jetted
but off-axis GRB orphan is not an easy task. The major problem is that
the trigger time is unknown. Some possible solutions to the problem 
are suggested.
\end{abstract}

\section {Introduction}

The detection of X-ray, optical and radio afterglows from some well-localized 
gamma-ray bursts (GRBs) definitely shows that at least most long GRBs are of 
cosmological origin (e.g., Costa et al. 1997; Frail et al. 1997;
Galama et al. 1998; Akerlof et al. 1999;  
Zhu et al. 1999). The so called fireball model is thus strongly favoured.  
However, we are still far from resolving the puzzle of GRBs (Piran 1999; van
Paradijs, Kouveliotou \& Wijers 2000). A major problem is that we do not know
whether GRBs are due to highly collimated jets or isotropic fireballs, so that
the energetics involved cannot be determined definitely 
(e.g., Pugliese, Falcke \& Biermann 1999; 
Kumar \& Piran 2000; Dar \& R\'{u}jula 2000; Wang \& Loeb 2001;
Rossi, Lazzati \& Rees 2002).
This issue has been extensively discussed in the literature. 

Generally speaking, three methods may help to determine the degree of 
beaming in GRBs. First, based on analytic solutions, it has been proposed 
that optical afterglows from a jetted GRB should be characterized by a 
break in the light curve during the relativistic phase, i.e., at the time 
when the Lorentz factor of the blastwave is $\gamma \sim 1/ \theta$, where 
$\theta$ is the half opening angle (Rhoads 1997; Kulkarni et al. 1999; 
M\'{e}sz\'{a}ros \& Rees 1999). Some GRBs such as 
990123, 990510 are regarded as good examples (Castro-Tirado et al. 1999; 
Wijers et al. 1999; Harrison et al. 1999; Halpern et al. 2000; Pian et al.
2001; Castro Ceron et al. 2001; Sagar et al. 2001). 
However, detailed numerical studies 
show that the break is usually quite smooth (Panaitescu \& M\'{e}sz\'{a}ros 
1998; Moderski, Sikora \& Bulik 2000), and Huang et al. went further to 
suggest that the light curve break should in fact occur at the 
trans-relativistic phase (Huang et al. 2000a, b, c, d). Additionally, many 
other factors can also result in light curve breaks, for example, the cooling 
of electrons (Sari, Piran \& Narayan 1998), a dense interstellar medium 
(Dai \&  Lu 1999), or a wind environment (Dai \& Lu 1998a; Chevalier 
\& Li 2000; Panaitescu \& Kumar 2000). All these facts combine together to 
make the first method not so conclusive. Second, Gruzinov (1999) argued that 
optical afterglows from a jet can be strongly polarized, in principle up to 
tens of percents. Some positive observations have already been reported 
(Wijers et al. 1999; Rol et al. 2000). But polarization 
can be observed only under some particular conditions, i.e., the co-moving 
magnetic fields parallel and perpendicular to the jet should have different 
strengths and we should observe at the right time from the right viewing 
angle (Gruzinov 1999; Hjorth et al. 1999; Mitra 2000). 

The third method was first proposed by Rhoads (1997), who pointed out that 
due to relativistic beaming effects, $\gamma$-ray radiation from the vast 
majority of jetted GRBs cannot be observed, but the corresponding late time 
afterglow emission is less beamed and can safely reach us. These afterglows 
are called orphan afterglows, which means they are not associated with any
detectable GRBs. The ratio of the orphan afterglow rate to the GRB rate 
might allow measurement of the GRB collimation angle. Great expectations have
been put on this method (Rhoads 1997; M\'{e}sz\'{a}ros, Rees \& Wijers 1999;
Lamb 2000; Paczy\'{n}ski 2000; Djorgovski et al. 2001). In fact, the
absence of large numbers of orphan afterglows in many surveys has been 
regarded as evidence 
that the collimation cannot be extreme (Rhoads 1997; Perna \& Loeb 1998;  
Greiner et al. 1999; Grindlay 1999; Rees 1999).

Recently Dalal, Griest \& Pruet (2002) argued that measurement of the GRB 
beaming angle using optical orphan searches is extremely difficult. The 
main reason is that when the afterglow emission from a jet begins to go 
into a much larger solid angle than the initial burst does, the optical 
flux density usually becomes very low. Generally speaking, this problem 
can be overcome by improving the detection limit.  In fact, an 
interesting result was recently reported by Vanden Berk et al. (2002),
who discovered a possible optical orphan at $z = 0.385$. They suggested 
that GRBs should be collimated. 

In this article, we will point out another difficulty associated 
with the third method: there should be many ``failed gamma-ray bursts 
(FGRBs)'', i.e., baryon-contaminated fireballs with initial Lorentz factor 
$\gamma_0 \ll 100$. FGRBs cannot be observed in gamma-rays, but their
long-lasting afterglows are detectable, thus they will also manifest
themselves as orphan afterglows. Our paper is organized as follows. 
In Section 2 we explain the concept of FGRBs. Section 3 describes 
the difficulty in distinguishing an FGRB orphan and a jetted GRB orphan.  
Some possible solutions that may help to overcome the difficulty are 
suggested. Section 4 is a brief discussion. 

\section{Failed Gamma-Ray Bursts and Their Afterglows}

Occurring in the deep Universe, GRBs are the most relativistic phenomena 
ever known. The standard fireball model (M\'{e}sz\'{a}ros \& Rees 1992;
Dermer, B\"{o}ttcher \& Chiang 1999) requires 
that to successfully produce a GRB, 
the initial Lorentz factor of the blastwave should typically be 
$\gamma_0 \geq 100$ --- 1000 during the main burst phase (Piran 1999;
Lithwick \& Sari 2000). Generally speaking 
the requirement of ultra-relativistic motion is to avoid the so called 
``compactness problem''. A modest variation in the Lorentz 
factor will result in a difference of the opacity of the high-energy 
$\gamma$-ray photons by a factor of $\sim 10^3$ (Totani 1999). 
Additionally, assuming synchrotron radiation, the observed peak 
frequency is strongly dependent on $\gamma$,
$\nu_{\rm m} \propto \gamma^4$ (M\'{e}sz\'{a}ros, Rees \& Wijers 1998). 
Thus a Lorentz factor of $\gamma_0 \leq 50$ makes the 
blastwave very inefficient in emitting $\gamma$-ray photons. 

So, to successfully produce a GRB, we need $\gamma_0 \geq 100$ --- 1000. 
However, theoretically it is not easy to construct a model to 
generate such ultra-relativistic motions. Currently there are mainly 
two kinds of progenitor models, the collapse of massive stars (with 
mass $M \geq 40 M_{\odot}$), or the collision of two compact stars (such
as two neutron stars or a neutron star and a black hole). 
Since a baryon-rich environment is involved in all these models,
some researchers are afraid that the baryon-contamination problem 
may exist. But this problem 
maybe is not as serious as we previously expected. 

Let us first take the collapsar model (MacFadyen \& Woosley 1999) as 
an example. We can imagine that the baryon mass and  
energy released in different collapsar events should 
vary greatly, then $\gamma_0$ of the resultant fireballs 
may also vary in a relatively wide range.  
In most cases, $\gamma_0$ should be very low (i.e.,
$\gamma_0 \ll 100$), but there still could be 
a few cases (e.g., one percent or even one in a thousand) in which
the fireball is relatively clean so that the blastwave can be 
successfully accelerated to $\gamma_0 \geq 100$ --- 1000 and produces
a GRB. Since the collapsar rate is high enough in a typical galaxy, 
there should be no problem that such collapsars can meet the 
requirement of GRB rate (i.e., $10^{-7}$ --- $10^{-6}$ event per 
typical galaxy per year, under isotropic assumption). Cases are 
similar in the collisions of two compact stars. 

In short, we cannot omit an important fact: if GRBs are really due 
to isotropic fireballs, then there should be much more failed GRBs
(i.e., fireballs with Lorentz factor much less than one hundred, but
still much greater than unity). These FGRB fireballs can contain 
similar initial energy as normal GRB fireballs, i.e., $E_0 \sim 10^{51}$
--- $10^{53}$ ergs, but they are polluted by baryons with mass 
$M_0 \sim 10^{-5}$ --- $10^{-3} M_{\odot}$. Radiation from these 
FGRBs should mainly be in x-ray bands in the initial bursting phase, 
not in $\gamma$-ray bands.  In fact, BeppoSAX team has reported 
the discovery of several anomalous events named as fast X-ray
transients, X-ray rich GRBs, or even X-ray-GRBs. They resemble 
usual GRBs except that they are extremely X-ray rich (Frontera 
et al. 2000; Kippen et al. 2001; Gandolfi \& Piro 2001). 
Observational data on this kind of events are being accumulated
rapidly. Recent good examples include GRBs 011030, 011130 and 
011211 (Gandolfi et al. 2001a, b; Ricker et al. 2001).   
We propose that these events are probably just FGRBs. 

Huang et al. (1998) and
Dai, Huang \& Lu (1999) have pointed out that for afterglow behaviour,
the parameter $E_0$ is decisive, while $M_0$ is only of minor 
importance, especially at late stages. So, FGRBs should also be 
associated with prominent afterglows. In Figure~1 we compare the 
theoretical optical afterglows from FGRBs with those from isotropic 
GRBs and jetted GRBs. We can see that the light curve of an FGRB 
afterglow differs from that of a successful isotropic burst only 
slightly, i.e., only notable at early stages. In our calculations, 
we have used the methods developed by Huang et al.(1999a, b, 
2000b, d), i.e., for 
the dynamical evolution of isotropic fireballs we use 
\begin{equation}
\label{dgdmass1}
\frac{{\rm d} \gamma}{{\rm d} m} = - \frac{\gamma^2 - 1}
       {M_0 + \epsilon m + 2 ( 1 - \epsilon) \gamma m},
\end{equation}
where $m$ is the swept-up mass and $\epsilon$ is the radiation efficience.
Eq. (1) has been proved to be proper in both ultra-relativistic phase
and non-relativistic phase (Huang, Dai \& Lu 1999a, b).
For jetted ejecta, the following equation is added (Huang et al. 2000b, d),
\begin{equation}
\label{dthetadt2}
\frac{{\rm d} \theta}{{\rm d} t} = 
\frac{c_{\rm s} (\gamma + \sqrt{\gamma^2 - 1})}{R},
\end{equation}
where $R$ is the blastwave radius, and the co-moving sound 
speed $c_{\rm s}$ is given realistically by 
\begin{equation}
\label{cssquare3}
c_{\rm s}^2 = \hat{\gamma} (\hat{\gamma} - 1) (\gamma - 1)
              \frac{1}{1 + \hat{\gamma}(\gamma - 1)} c^2, 
\end{equation}
with $\hat{\gamma} = (4 \gamma + 1) / (3 \gamma)$ the adiabatic index.

In fact, in beamed GRB models, there should also be many FGRBs, i.e., 
beamed ejecta with $1 \ll \gamma_0 \ll 100$. We call them beamed FGRBs. 
Afterglow from beamed FGRBs has also been illustrated in Figure~1. In  
this article emphasises will be put on isotropic FGRBs, so by using 
``FGRBs'' we will only mean isotropic FGRBs unless stated explicitly. 

\section{Orphan Afterglows}

Both FGRBs and jetted but off-axis GRBs can produce isolated fading 
objects, i.e., orphan afterglows. Theoretically, when orphan afterglows 
are really discovered observationally, it is still risky to conclude that 
GRBs are beamed. We should study these orphans carefully to determine 
whether they come from FGRBs or Jetted GRBs. 
However, we will show below that it is not an easy task.

\subsection{Difficulty in using orphan afterglows}

Usually the light curve of GRB afterglows is plotted as $\log S_{\rm \nu}$
vs. $\log t$, where $S_{\rm \nu}$ is the flux density at observing 
frequency $\nu$ and $t$ is observer's time measured from the burst 
trigger. In such plots, the behaviour of afterglows from isotropic GRBs and jetted 
ones are possibly quite different. The former is generally characterized 
by a simple flat straight line with slope $ \sim -1.0$ --- $-1.3$ and the 
latter can be characterized by a break in the light curve or by
a steep straight line with a slope sharper than $\sim -2.0$ (Figure~1,
also see Huang et al. 2000a, b, c, d). 
But for orphan afterglow observations, the derivation of such a 
$\log S_{\rm \nu}$ --- $\log t$ light curve is not direct: we 
do not know the trigger time so that the exact value of $t$ for each 
observed data point cannot be determined. 

As the first step,
the best that we can do is to produce a light curve with a linear
time axis, which, however, is of little help for unveiling the nature of 
the orphan. Figure 2 illustrates the matter. In this figure we  
plot $\log S_{\rm \nu}$ vs. $t$ for the two kinds of orphans theoretically.   
The uncerntainty in trigger time means the observed light curve can be shifted 
along X axis, while the unknown distance results in a shift along Y axis. 
We see that after some simple manipulations, the segment AB on the dashed 
curve (i.e., from $t = 18$ d to $t = 55$ d) 
can be shifted to a place (A$'$B$'$) 
that differs from the solid line only slightly. 
It hints that a linear light curve as long as $\sim 37$ days is still not enough. 
Note that the solid and the dashed curves in Figure 2 are only two examples.
The variation of some intrinsic parameters, such as $E_0$, $n$, $\gamma_0$, 
$\xi_{\rm e}$, $\xi_{\rm B}^2$, $p$, $\theta_0$ and $\theta_{\rm obs}$ 
as defined in the caption of Figure 1,
can change the shape of the two curves notably, thus brings in much more
difficulties. 

In Figure 3, we compare the theoretical $\log S_{\rm \nu}$ --- $\log t$ 
light curves of optical afterglows 
from FGRBs and jetted but off-axis GRBs directly. To investigate the 
influence of the uncertainty in trigger time, we also shift the 
light curve of FGRBs by $t \pm 3$ d, $t \pm 10$ d and $t \pm 30$ d 
intentionally. From the dashed curves, we can see that the shape of 
the FGRB afterglow light curve is seriously affected by the uncertainty 
of the trigger time. But fortunately, these dashed curves still 
differ from the theoretical light curve of the jetted GRB orphan 
markedly, i.e., they are much flatter at very late stages. This means
it is still possible for us to discriminate them. 
In Figure~4, similar results to Figure~3 are given, but this 
time the light curve of the jetted GRB orphan is shifted. Again we see
that the two kinds of orphans can be discriminated by their late time 
behaviour. 

Figures 3 and 4 explain what we should do when an orphan afterglow is 
discovered. First, we have to assume a trigger time for it arbitrarily, 
so that the logarithmic 
light curve can be plotted. We then need to change the 
trigger time to many other values to see how the light curve is 
affected. In all our plots, we should pay special attention to the 
late time behaviour, which will be less affected by the uncertainty in 
the trigger time. If the slope tends to be $\sim -1.0$ --- $-1.3$, then
the orphan afterglow may come from an FGRB event. But if the slope 
tends to be steeper than $\sim -2.0$, then it is very likely from 
a jetted but off-axis GRB. 

In fact, from Figure 1, we know that for all kinds of GRBs, either
successful or failed, the optical afterglow approximately follows 
a simple power-law decay at late stages (i.e., 
$S_{\rm R} \approx k t^{-\alpha}$) so that the light curve is
a straight line. In such a relation, if we shift the time by $T$,  
\begin{equation}
\label{SRT}
S_{\rm R} \approx k (t+T)^{-\alpha},
\end{equation}
then the line would become curved. The slope at each point on the
curve is 
\begin{equation}
\label{slope}
\frac{{\rm d} \log S_{\rm R}}{{\rm d} \log t} = 
    - \alpha ( \frac{t}{t+T} ).
\end{equation}
For positive $T$ values the lines bend up-ward, while for 
negative values the lines bend down-ward. It hints 
that in plotting the orphan afterglow light curve, we could 
select the trigger time properly to get a straight line 
at late stages, then we can determine not only the late
time slope, but also the true trigger time.   
In other words, we can use
\begin{equation}
\label{slope2}
\frac{{\rm d}^2 \log S_{\rm R}}{{\rm d} (\log t)^2} \equiv 0
\end{equation}
as the condition to determine the trigger time and to get
the straight line at late stages.

However, we must bear in mind that it is in fact not an easy 
task. First, to take the process we need 
to follow the orphan as long as possible, and 
the simple discovery of an orphan is obviously insufficient. 
Note that currently optical afterglows from most 
well-localized GRBs can be observed for only less than 100 days. 
It is quite unlikely that we can follow an orphan for 
a period longer than that.  
Second, since the orphan is usually very faint, 
errors in the measured 
magnitudes will seriously prevent us from deriving the straight  
line. Due to all these difficulties, a satisfactory light curve
is usually hard to get for most orphans. 

We see that measurement of the GRB 
beaming angle using orphan searches is not as simple as we 
originally expected. In fact, it is impractical to some extent.  
Recently it was suggested by Rhoads (2001) that GRB afterglows 
can be effectly identified by snapshot observations made with 
three or more optical filters. The method has been successfully 
applied to GRB 001011 by Gorosabel et al. (2001). It is believed 
that this method is also helpful for orphan afterglow searches. 
However, please note that a jetted GRB orphan and an FGRB one 
still cannot be discriminated directly. 

\subsection{Possible solutions}

We have shown that the derivation of a satisfactory light curve
for an orphan afterglow is difficult. The major problem is that
we do not know the trigger time. Anyway, there are
still some possible solutions that may help to determine 
the onset of an orphan afterglow.  

Firstly, of course we should improve our detection limit so 
that the orphan afterglow could be followed as long as possible.
The longer we observe, the more likely that we can get the true
late-time light curve slope. Secondly, we know that FGRBs usually 
manifest themselves as fast X-ray transients or X-ray-GRBs. If 
an orphan can be identified to associate with such a transient, 
then it is most likely an FGRB one. In this case, the trigger 
time can be well determined.  

Thirdly, maybe in some rare cases we are 
so lucky that the rising phase of the orphan could be observed.
For a jetted GRB orphan the maximum optical flux is usually 
reached within one or two days and for an FGRB orphan it is even
within hours. Then the uncertainty in trigger time is greatly 
reduced. Additionally, a jetted GRB orphan differs markedly 
from an FGRB one during the rising phase. The former can be 
brightened by more than one magnitude in several hours (see 
Figures~1 --- 4), while the brightening of the latter can hardly 
be observed. So, if an orphan afterglow with a short period of 
brightening is observed, then it is most likely a jetted GRB 
orphan. Of course, we should first be certain that it is not 
a supernova. 

Fourthly, valuable clues may come from radio observations. In radio
bands, the light curve should be highly variable at early stages 
due to interstellar medium scintillation, and it will become much
smoother at late times. So the variability in radio light curves 
provides useful information on the trigger time. And fifthly, in 
the future maybe gravitational wave radiation 
or neutrino radiation associated with GRBs
could be detected due to progresses in technology, then the trigger
time of an orphan could be determined directly and accurately. In 
fact, with the successful detection of gravitational waves
or neutrino emission, our understanding on GRB progenitors will 
surely be promoted greatly (Paolis et al. 2001). 

Sixthly, the redshift of the orphan afterglow can help us greatly
in determining the isotropic energy involved, which itself is 
helpful for inferring the trigger time. Seventhly, the microlensing 
effect may be of some help. Since the size of the radiation zone
of a jetted GRB orphan is much smaller than that of an FGRB one, 
they should behave differently when microlensed. 

Finally, although a successful detection of some orphan afterglows
does not directly mean that GRBs be collimated, the negative 
detection of any orphans can always place both a stringent lower limit
on the beaming angle for GRBs and a reasonable upper limit for the 
rate of FGRBs.  

\section{Conclusion and Discussion}

To successfully produce a GRB, the blastwave should be
ultra-relativistic, with Lorentz factor typically larger than 100
--- 1000. However, in almost all popular progenitor models, the
environment is unavoidably baryon-rich. We believe that only in very
rare cases can an ultra-relativistic blastwave successfully break 
out to give birth to a GRB, and there should be much more failed
GRBs, i.e., fireballs with Lorentz factor much less than 100 but 
still much larger than unity. In fact, this possibility has also 
been mentioned by a number of authors, such as M\'{e}sz\'{a}ros 
\& Waxman (2001). 

Owing to the existence of FGRBs, there should be many orphan 
afterglows even if GRBs are due to isotropic fireballs. Then the 
simple discovery of orphan afterglows does not necessarily mean 
that GRBs be highly collimated. To make use of information from
orphan afterglow surveys correctly, we should first know how to 
discriminate a jetted GRB orphan and an FGRB one. This can 
be done only by checking the detailed afterglow light curve. 
However, we have shown that the derivation of a satisfactory 
light curve for an orphan afterglow is difficult. The major 
problem is that we do not know the trigger time. 

In Section 3.2, some possible solutions to the problem are 
suggested. Unfortunately many of these solutions are still 
quite impractical in the foreseeable future, which means 
measure of GRB beaming angle using orphan afterglow searches 
is extremely difficult currently. However, special attention 
should be paid to the second solution. Usually, FGRBs  
manifested themselves as fast X-ray transients 
during the main burst phase,
while jetted but off-axis GRBs went unattended completely.   
If the fast X-ray transients (or X-ray-GRBs) observed by 
BeppoSAX are really due to FGRBs, then afterglows should be
detectable. We propose that this kind of events should be 
followed rapidly and extensively in all bands, just like what 
we are doing for GRBs. If observed, afterglows from these 
anomalous events can be used to check our concept of FGRBs,  
and even to test the fireball model under quite different
conditions (i.e., when $\gamma_0 \ll 100$). Also, these FGRBs 
can provide valuable information for our understanding of GRBs, 
especially on the progenitor models. 
Note that beamed FGRBs can also give birth to fast X-ray 
transients if they are directed toward us, but afterglows 
from such a beamed FGRB and an isotropic FGRB can be 
discriminated easily from the light curves (see Figure~1). 

It is very interesting to note that optical afterglows from
two X-ray-GRBs, 011130 and 011211, have been observed 
(Garnavich, Jha \& Kirshner 2001; Grav et al. 2001).
Their redshifts were measured to be $z = $ 0.5 and 2.14 
respectively (Jha et al. 2001; Fruchter et al. 2001), 
eliminates the possibility that they were
ordinary classic GRBs residing at extremely high redshifts
($z \geq 10$). We propose that they should be FGRBs 
(either isotropic or beamed) or just jetted GRB ``orphan''.
However, the observational data currently available are still
quite poor so that we could not determine their nature definitely. 
As for other X-ray-GRBs without a measured 
redshift, the possibility that they were at redshifts of 
$z \geq 10 $ can not be excluded. 

Finally, the concept of FGRBs is based on the fact that most
popular progenitor models for GRBs are baryon-rich. But cases are
quite different for another kind of progenitor models where
strange stars are involved. Strange stars, composed mainly of u, 
d, and s quarks, are compact objects which are quite similar to 
neutron stars observationally (Alcock, Farhi \& Olinto 1986). 
A typical strange star (with mass $\sim 1.4 M_\odot$) can have 
a normal matter crust of less than $\sim 2 \times 10^{-5} M_{\odot}$ 
(Alcock, Farhi \& Olinto 1986), or
even as small as $\sim 3 \times 10^{-6} M_{\odot}$ 
(Huang \& Lu 1997a, b). Then baryon contamination can be 
directly avoided if GRBs are due to the phase transition of 
neutron stars to strange stars (Cheng \& Dai 1996;
Dai \& Lu 1998b) or 
collisions of binary strange stars. 
In these models, there should be very few FGRBs.   

\section*{Acknowledgments}
We thank an anonymous referee for valuable comments and suggestions.
YFH thanks L. J. Gou and X. Y. Wang for helpful discussion. 
This research was supported by the Special Funds for Major State 
Basic Research Projects, the National Natural Science Foundation 
of China (grants 10003001, 19825109, and 19973003), and the 
National 973 Project (NKBRSF G19990754). 

Note added after acceptance {\bf (this paragraph might not appear in 
the published version)}: The optical orphan at z=0.385 reported by 
Vanden Berk et al. (2002) has recently been proved to be an 
unusual radio-loud AGN (Gal-yam et al. 2002, astro-ph/0202354), 
and X-ray GRB 011211 was found to be in fact an ordinary 
classic GRB (Frontera et al., GCN 1215). Additionally, the 
optical identification of X-ray GRB 011130 might also be 
incorrect (Frail et al., GCN 1207). We sincerely thank 
Nicola Masetti for private communication. 

{}

\clearpage

\begin{figure} \centering 
\epsfig{file=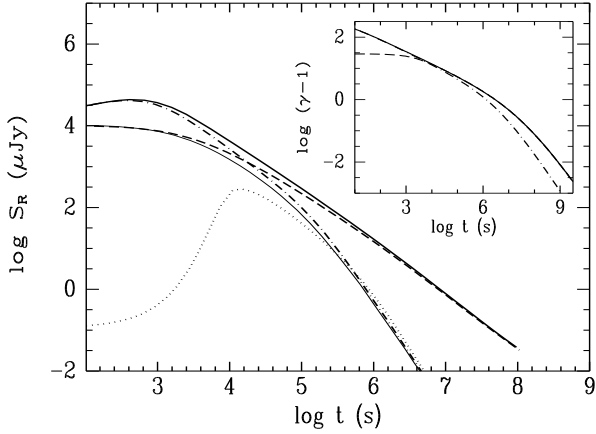, angle=-90, height=60mm, width=6.8cm, 
bbllx=120pt, bblly=125pt, bburx=530pt, bbury=575pt}
\caption{ Theoretical R band light curves of GRB afterglows. In 
  our calculations we have taken the following common parameters:
  the isotropic energy $E_0 =10^{53}$ ergs, the interstellar medium
  density $n = 1$ cm$^{-3}$, electron energy fraction $\xi_{\rm e}
  = 0.1$, magnetical energy fraction $\xi_{\rm B}^2 = 10^{-4}$,
  electron power-law energy index $p=2.5$, and the luminosity 
  distance $d = 1$ Gpc. For jets, we take the initial half opening
  angle $\theta_0 = 0.1$. The thick solid line is plotted for a usual 
  isotropic GRB with $\gamma_0 = 300$. The dashed line represents
  an isotropic FGRB orphan with $\gamma_0 =30$. The dash-dotted line 
  corresponds to an on-axis jetted GRB with $\gamma_0 = 300$, and 
  the dotted line is for a jetted but off-axis GRB orphan with 
  viewing angle $\theta_{\rm obs} = 0.15$. The thin solid line is for
  a beamed FGRB with $\gamma_0 = 30, \theta_{\rm obs} =0$. Inset shows
  the evolution of the Lorentz factor correspondingly. Note that 
  $\gamma (t)$ of the beamed FGRB is not shown, since it is too 
  close to the dashed curve at early times and too close to the 
  dash-dotted line at late stages. } 
\label{fig1}
\end{figure}

\begin{figure} \centering
\epsfig{file=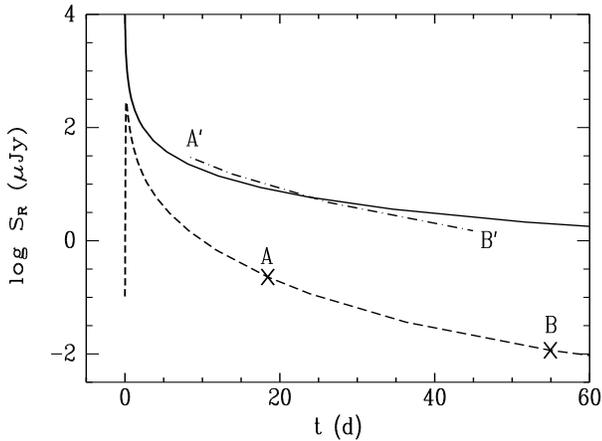, angle=-90, height=60mm, width=6.8cm, 
bbllx=120pt, bblly=125pt, bburx=530pt, bbury=575pt}
\caption{ Sample light curves of the two kinds of orphan afterglows 
  plotted with a linear time axis.  
  The solid line represents an isotropic
  FGRB and the dashed line represents a 
  jetted but off-axis GRB orphan. 
  Parameters are the same as in Fig. 1. 
  Points A and B on the dashed 
  curve are at $t = 18$ d and $t= 55$ d respectively.   
  The dash-dotted line (A$'$B$'$) is plotted by shifting  
  curve AB with $t - 10$ d and $S_{\rm R} \times 130$.
  Note that A$'$B$'$ deviates from the solid line only
  marginally. }   
\label{fig2}
\end{figure}

\begin{figure} \centering
\epsfig{file=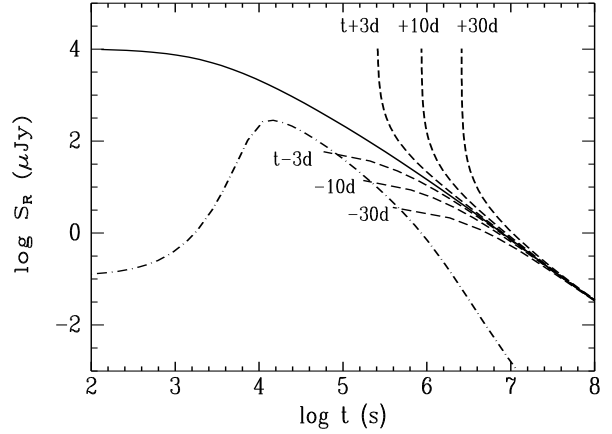, angle=-90, height=60mm, width=6.8cm, 
bbllx=120pt, bblly=125pt, bburx=530pt, bbury=575pt}
\caption{ Direct comparison of the two kinds of orphan afterglows. 
  The solid line represents the optical light curve of an isotropic
  FGRB and the dash-dotted line corresponds to a 
  jetted but off-axis GRB orphan. The 
  dashed lines are drawn by shifting the solid line by 
  $t \pm 3$ d, $t \pm 10$ d, and $t \pm 30$ d respectively. 
  Parameters are the same as in Fig. 1.} 
\label{fig3}
\end{figure}

\begin{figure} \centering 
\epsfig{file=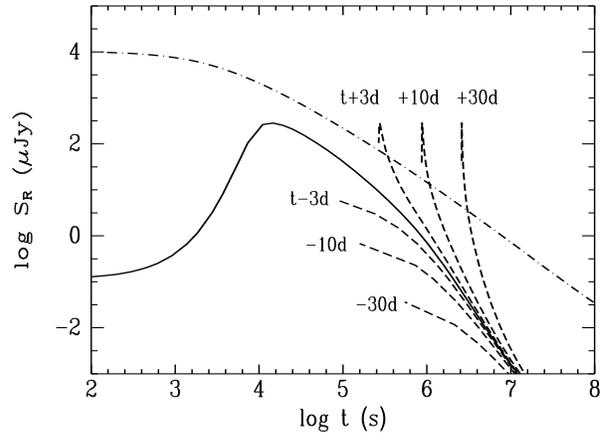, angle=-90, height=60mm, width=6.8cm, 
bbllx=120pt, bblly=125pt, bburx=530pt, bbury=575pt}
\caption{ The same as in Fig. 3, but this time the light curve of 
the jetted GRB orphan is shifted.}
\label{fig4}
\end{figure}

\end{document}